\documentclass[%
reprint,
superscriptaddress,
showpacs,
 amsmath,amssymb,
 aps,
]{revtex4-1}

\usepackage{graphicx}
\usepackage{dcolumn}
\usepackage{bm}
\usepackage{amsmath}
\usepackage{autobreak}
\usepackage{hyperref}
\usepackage[mathlines]{lineno}
\usepackage{natbib}
\usepackage{multirow}
\usepackage{booktabs}
\usepackage{xspace}
\usepackage{float}
\usepackage{diagbox}

\begin{document}

\preprint{APS/123-QED}

\title{Probing Dark Matter Particles from Evaporating Primordial Black Holes via Electron Scattering in the CDEX-10 Experiment}

\author{Z.~H.~Zhang}
\affiliation{Key Laboratory of Particle and Radiation Imaging (Ministry of Education) and Department of Engineering Physics, Tsinghua University, Beijing 100084}
\author{L.~T.~Yang}
\email{Corresponding author: yanglt@mail.tsinghua.edu.cn}
\affiliation{Key Laboratory of Particle and Radiation Imaging (Ministry of Education) and Department of Engineering Physics, Tsinghua University, Beijing 100084}
\author{Q.~Yue}
\email{Corresponding author: yueq@mail.tsinghua.edu.cn}
\affiliation{Key Laboratory of Particle and Radiation Imaging (Ministry of Education) and Department of Engineering Physics, Tsinghua University, Beijing 100084}
\author{K.~J.~Kang}
\affiliation{Key Laboratory of Particle and Radiation Imaging (Ministry of Education) and Department of Engineering Physics, Tsinghua University, Beijing 100084}
\author{Y.~J.~Li}
\affiliation{Key Laboratory of Particle and Radiation Imaging (Ministry of Education) and Department of Engineering Physics, Tsinghua University, Beijing 100084}

\author{H.~P.~An}
\affiliation{Key Laboratory of Particle and Radiation Imaging (Ministry of Education) and Department of Engineering Physics, Tsinghua University, Beijing 100084}
\affiliation{Department of Physics, Tsinghua University, Beijing 100084}

\author{Greeshma~C.}
\altaffiliation{Participating as a member of TEXONO Collaboration}
\affiliation{Institute of Physics, Academia Sinica, Taipei 11529}

\author{J.~P.~Chang}
\affiliation{NUCTECH Company, Beijing 100084}

\author{Y.~H.~Chen}
\affiliation{YaLong River Hydropower Development Company, Chengdu 610051}
\author{J.~P.~Cheng}
\affiliation{Key Laboratory of Particle and Radiation Imaging (Ministry of Education) and Department of Engineering Physics, Tsinghua University, Beijing 100084}
\affiliation{College of Nuclear Science and Technology, Beijing Normal University, Beijing 100875}
\author{W.~H.~Dai}
\affiliation{Key Laboratory of Particle and Radiation Imaging (Ministry of Education) and Department of Engineering Physics, Tsinghua University, Beijing 100084}
\author{Z.~Deng}
\affiliation{Key Laboratory of Particle and Radiation Imaging (Ministry of Education) and Department of Engineering Physics, Tsinghua University, Beijing 100084}
\author{C.~H.~Fang}
\affiliation{College of Physics, Sichuan University, Chengdu 610065}
\author{X.~P.~Geng}
\affiliation{Key Laboratory of Particle and Radiation Imaging (Ministry of Education) and Department of Engineering Physics, Tsinghua University, Beijing 100084}
\author{H.~Gong}
\affiliation{Key Laboratory of Particle and Radiation Imaging (Ministry of Education) and Department of Engineering Physics, Tsinghua University, Beijing 100084}
\author{Q.~J.~Guo}
\affiliation{School of Physics, Peking University, Beijing 100871}
\author{T.~Guo}
\affiliation{Key Laboratory of Particle and Radiation Imaging (Ministry of Education) and Department of Engineering Physics, Tsinghua University, Beijing 100084}
\author{X.~Y.~Guo}
\affiliation{YaLong River Hydropower Development Company, Chengdu 610051}
\author{L.~He}
\affiliation{NUCTECH Company, Beijing 100084}
\author{S.~M.~He}
\affiliation{YaLong River Hydropower Development Company, Chengdu 610051}
\author{J.~W.~Hu}
\affiliation{Key Laboratory of Particle and Radiation Imaging (Ministry of Education) and Department of Engineering Physics, Tsinghua University, Beijing 100084}
\author{H.~X.~Huang}
\affiliation{Department of Nuclear Physics, China Institute of Atomic Energy, Beijing 102413}
\author{T.~C.~Huang}
\affiliation{Sino-French Institute of Nuclear and Technology, Sun Yat-sen University, Zhuhai 519082}
\author{L.~Jiang}
\affiliation{Key Laboratory of Particle and Radiation Imaging (Ministry of Education) and Department of Engineering Physics, Tsinghua University, Beijing 100084}
\author{S.~Karmakar}
\altaffiliation{Participating as a member of TEXONO Collaboration}
\affiliation{Institute of Physics, Academia Sinica, Taipei 11529}

\author{H.~B.~Li}
\altaffiliation{Participating as a member of TEXONO Collaboration}
\affiliation{Institute of Physics, Academia Sinica, Taipei 11529}
\author{H.~Y.~Li}
\affiliation{College of Physics, Sichuan University, Chengdu 610065}
\author{J.~M.~Li}
\affiliation{Key Laboratory of Particle and Radiation Imaging (Ministry of Education) and Department of Engineering Physics, Tsinghua University, Beijing 100084}
\author{J.~Li}
\affiliation{Key Laboratory of Particle and Radiation Imaging (Ministry of Education) and Department of Engineering Physics, Tsinghua University, Beijing 100084}
\author{Q.~Y.~Li}
\affiliation{College of Physics, Sichuan University, Chengdu 610065}
\author{R.~M.~J.~Li}
\affiliation{College of Physics, Sichuan University, Chengdu 610065}
\author{X.~Q.~Li}
\affiliation{School of Physics, Nankai University, Tianjin 300071}
\author{Y.~L.~Li}
\affiliation{Key Laboratory of Particle and Radiation Imaging (Ministry of Education) and Department of Engineering Physics, Tsinghua University, Beijing 100084}
\author{Y.~F.~Liang}
\affiliation{Key Laboratory of Particle and Radiation Imaging (Ministry of Education) and Department of Engineering Physics, Tsinghua University, Beijing 100084}
\author{B.~Liao}
\affiliation{College of Nuclear Science and Technology, Beijing Normal University, Beijing 100875}
\author{F.~K.~Lin}
\altaffiliation{Participating as a member of TEXONO Collaboration}
\affiliation{Institute of Physics, Academia Sinica, Taipei 11529}
\author{S.~T.~Lin}
\affiliation{College of Physics, Sichuan University, Chengdu 610065}
\author{J.~X.~Liu}
\affiliation{Key Laboratory of Particle and Radiation Imaging (Ministry of Education) and Department of Engineering Physics, Tsinghua University, Beijing 100084}
\author{S.~K.~Liu}
\affiliation{College of Physics, Sichuan University, Chengdu 610065}
\author{Y.~D.~Liu}
\affiliation{College of Nuclear Science and Technology, Beijing Normal University, Beijing 100875}
\author{Y.~Liu}
\affiliation{College of Physics, Sichuan University, Chengdu 610065}
\author{Y.~Y.~Liu}
\affiliation{College of Nuclear Science and Technology, Beijing Normal University, Beijing 100875}
\author{H.~Ma}
\affiliation{Key Laboratory of Particle and Radiation Imaging (Ministry of Education) and Department of Engineering Physics, Tsinghua University, Beijing 100084}
\author{Y.~C.~Mao}
\affiliation{School of Physics, Peking University, Beijing 100871}
\author{Q.~Y.~Nie}
\affiliation{Key Laboratory of Particle and Radiation Imaging (Ministry of Education) and Department of Engineering Physics, Tsinghua University, Beijing 100084}
\author{J.~H.~Ning}
\affiliation{YaLong River Hydropower Development Company, Chengdu 610051}
\author{H.~Pan}
\affiliation{NUCTECH Company, Beijing 100084}
\author{N.~C.~Qi}
\affiliation{YaLong River Hydropower Development Company, Chengdu 610051}
\author{J.~Ren}
\affiliation{Department of Nuclear Physics, China Institute of Atomic Energy, Beijing 102413}
\author{X.~C.~Ruan}
\affiliation{Department of Nuclear Physics, China Institute of Atomic Energy, Beijing 102413}
\author{M.~K.~Singh}
\altaffiliation{Participating as a member of TEXONO Collaboration}
\affiliation{Institute of Physics, Academia Sinica, Taipei 11529}
\affiliation{Department of Physics, Banaras Hindu University, Varanasi 221005}
\author{T.~X.~Sun}
\affiliation{College of Nuclear Science and Technology, Beijing Normal University, Beijing 100875}
\author{C.~J.~Tang}
\affiliation{College of Physics, Sichuan University, Chengdu 610065}
\author{Y.~Tian}
\affiliation{Key Laboratory of Particle and Radiation Imaging (Ministry of Education) and Department of Engineering Physics, Tsinghua University, Beijing 100084}
\author{G.~F.~Wang}
\affiliation{College of Nuclear Science and Technology, Beijing Normal University, Beijing 100875}
\author{J.~Z.~Wang}
\affiliation{Key Laboratory of Particle and Radiation Imaging (Ministry of Education) and Department of Engineering Physics, Tsinghua University, Beijing 100084}
\author{L.~Wang}
\affiliation{Department of  Physics, Beijing Normal University, Beijing 100875}
\author{Q.~Wang}
\affiliation{Key Laboratory of Particle and Radiation Imaging (Ministry of Education) and Department of Engineering Physics, Tsinghua University, Beijing 100084}
\affiliation{Department of Physics, Tsinghua University, Beijing 100084}
\author{Y.~F.~Wang}
\affiliation{Key Laboratory of Particle and Radiation Imaging (Ministry of Education) and Department of Engineering Physics, Tsinghua University, Beijing 100084}
\author{Y.~X.~Wang}
\affiliation{School of Physics, Peking University, Beijing 100871}
\author{H.~T.~Wong}
\altaffiliation{Participating as a member of TEXONO Collaboration}
\affiliation{Institute of Physics, Academia Sinica, Taipei 11529}
\author{S.~Y.~Wu}
\affiliation{YaLong River Hydropower Development Company, Chengdu 610051}
\author{Y.~C.~Wu}
\affiliation{Key Laboratory of Particle and Radiation Imaging (Ministry of Education) and Department of Engineering Physics, Tsinghua University, Beijing 100084}
\author{H.~Y.~Xing}
\affiliation{College of Physics, Sichuan University, Chengdu 610065}
\author{R. Xu}
\affiliation{Key Laboratory of Particle and Radiation Imaging (Ministry of Education) and Department of Engineering Physics, Tsinghua University, Beijing 100084}
\author{Y.~Xu}
\affiliation{School of Physics, Nankai University, Tianjin 300071}
\author{T.~Xue}
\affiliation{Key Laboratory of Particle and Radiation Imaging (Ministry of Education) and Department of Engineering Physics, Tsinghua University, Beijing 100084}
\author{Y.~L.~Yan}
\affiliation{College of Physics, Sichuan University, Chengdu 610065}
\author{N.~Yi}
\affiliation{Key Laboratory of Particle and Radiation Imaging (Ministry of Education) and Department of Engineering Physics, Tsinghua University, Beijing 100084}
\author{C.~X.~Yu}
\affiliation{School of Physics, Nankai University, Tianjin 300071}
\author{H.~J.~Yu}
\affiliation{NUCTECH Company, Beijing 100084}
\author{J.~F.~Yue}
\affiliation{YaLong River Hydropower Development Company, Chengdu 610051}
\author{M.~Zeng}
\affiliation{Key Laboratory of Particle and Radiation Imaging (Ministry of Education) and Department of Engineering Physics, Tsinghua University, Beijing 100084}
\author{Z.~Zeng}
\affiliation{Key Laboratory of Particle and Radiation Imaging (Ministry of Education) and Department of Engineering Physics, Tsinghua University, Beijing 100084}
\author{B.~T.~Zhang}
\affiliation{Key Laboratory of Particle and Radiation Imaging (Ministry of Education) and Department of Engineering Physics, Tsinghua University, Beijing 100084}
\author{F.~S.~Zhang}
\affiliation{College of Nuclear Science and Technology, Beijing Normal University, Beijing 100875}
\author{L.~Zhang}
\affiliation{College of Physics, Sichuan University, Chengdu 610065}
\author{Z.~Y.~Zhang}
\affiliation{Key Laboratory of Particle and Radiation Imaging (Ministry of Education) and Department of Engineering Physics, Tsinghua University, Beijing 100084}
\author{J.~Z.~Zhao}
\affiliation{Key Laboratory of Particle and Radiation Imaging (Ministry of Education) and Department of Engineering Physics, Tsinghua University, Beijing 100084}
\author{K.~K.~Zhao}
\affiliation{College of Physics, Sichuan University, Chengdu 610065}
\author{M.~G.~Zhao}
\affiliation{School of Physics, Nankai University, Tianjin 300071}
\author{J.~F.~Zhou}
\affiliation{YaLong River Hydropower Development Company, Chengdu 610051}
\author{Z.~Y.~Zhou}
\affiliation{Department of Nuclear Physics, China Institute of Atomic Energy, Beijing 102413}
\author{J.~J.~Zhu}
\affiliation{College of Physics, Sichuan University, Chengdu 610065}

\collaboration{CDEX Collaboration}
\noaffiliation

\date{\today}

\begin{abstract}
Dark matter (DM) is a major constituent of the Universe. However, no definite evidence of DM particles (denoted as ``$\chi$") has been found in DM direct detection (DD) experiments to date. There is a novel concept of detecting $\chi$ from evaporating primordial black holes (PBHs). We search for $\chi$ emitted from PBHs by investigating their interaction with target electrons. The examined PBH masses range from 1$\times$10$^{15}$ to 7$\times$10$^{16}$ g under the current limits of PBH abundance $f_{PBH}$. Using 205.4 kg$\cdot$day data obtained from the CDEX-10 experiment conducted in the China Jinping Underground Laboratory, we exclude the $\chi$--electron ($\chi$--$e$) elastic-scattering cross section $\sigma_{\chi e} \sim 5\times10^{-29}$ cm$^2$ for $\chi$ with a mass $m_{\chi}\lesssim$ 0.1 keV from our results. With the higher radiation background but lower energy threshold (160 eV), CDEX-10 fill a part of the gap in the previous work. If ($m_{\chi}$, $\sigma_{\chi e}$) can be determined in the future, DD experiments are expected to impose strong constraints on $f_{PBH}$ for large $M_{PBH}$s.
\end{abstract}

\maketitle

\section{Introduction}\label{sec:introduction}

There is strong evidence indicating that dark matter (DM) is a major constituent of the Universe~\cite{Bertone_2018, Workman_2022ynf,Matarrese_2010,Bertone_2005}. Direct detection (DD), indirect detection~\cite{DAMPE}, and collider~\cite{Daci_2015, DarkSHINE1, DarkSHINE2, DarkSHINE3} experiments have been conducted for decades to search for DM particles (denoted as ``$\chi$"). In DD experiments, the most popular search object is the $\chi$ in the Standard Halo Model (SHM). The velocity of $\chi$ in the SHM follows the Maxwell-Boltzmann distribution with a most probable value of 238 km/s and a cutoff at 544 km/s~\cite{VDis_2007, VDis_2021}. Time projection chambers (XENON~\cite{XENONnT_2023}, LUX-ZEPLIN~\cite{LZ_2023}, PandaX~\cite{PandaX-4T_2023}, DarkSide~\cite{Chi-e_DarkSide_2018}), cryogenic calorimeters (CRESST~\cite{CRESST_2019}, SuperCDMS~\cite{Chi-e_SuperCDMS_2020}, EDELWEISS~\cite{Chi-e_EDELWEISS_2020}), charge-coupled devices (SENSEI~\cite{Chi-e_SENSEI_2020}, DAMIC~\cite{Chi-e_DAMIC_2019}), and high purity germanium detectors (CoGeNT~\cite{CoGeNT_2013}, CDEX~\cite{CDEX_LSK_2014, CDEX_ZW_2013, CDEX_YQ_2014, CDEX_ZW_2016, CDEX_YLT_2018, CDEX_JH_2018, CDEX_JH_2019, CDEX_YLT_2019, CDEX_WY_2021, CDEX_GXP_2023, CDEX_ZZY_2022}) have been operated in searching for $\chi$ by investigating their interaction with target electrons or target nuclei.

In the paradigm of the $\chi$--nucleus ($\chi$--$N$) scattering, the best detection sensitivity has been achieved for $m_\chi\sim\mathcal{O}$(10 GeV)~\cite{LZ_2023}, where $m_\chi$ is the mass of $\chi$. As $m_\chi$ decreases, little energy transfers to the heavy nucleus in the elastic scattering process, resulting in a rapid decrease in the detection sensitivity. Considering the Migdal effect, the sensitivity in the low $m_\chi\sim\mathcal{O}$(100 MeV) range can be significantly improved~\cite{CDEX_LZZ_2019, Migdal_LUX_2019, Migdal_CRESST_2017, CDEX_LZZ_2022}. Further, when $\chi$ are scattered and accelerated to (semi)relativistic velocities by cosmic ray nuclei~\cite{CRDM_Cappiello_2019, CRDM_CDEX_2022, CRDM_Bringmann_2019, CRDM_Cappiello_063004, CRDM_Ema_2021, CRDM_Andriamirado_2021, CRDM_PandaX_2022}, blazar jet protons~\cite{BlazarBDM_Wang}, and supernova remnants~\cite{Suppernova_Cappiello}, $\chi$ with $m_\chi\lesssim\mathcal{O}$(1 GeV) can be searched.

In another paradigm parallel to $\chi$--$N$ scattering, the $\chi$--electron ($\chi$--$e$) scattering, the best detection sensitivity has been achieved for SHM $\chi$ with a mass of $m_\chi\sim\mathcal{O}$(10 MeV)~\cite{CDEX_ZZY_2022} rather than $m_\chi\sim\mathcal{O}(m_e)=\mathcal{O}$(0.5 MeV) because of the reduced $\chi$ kinetic energy. Similar to the $\chi - N$ scenario, $\chi$ can be accelerated by cosmic ray electrons~\cite{CRDM_Cappiello_2019, CRDM_Cappiello_063004, CRDM_Ema_2019}, cosmic ray neutrinos~\cite{CRDM_Wen_2019}, stellar neutrinos~\cite{stellar_neutrino1, stellar_neutrino2}, diffuse supernova neutrinos~\cite{supernova_neutrino1, supernova_neutrino2}, stellar electrons~\cite{SRDM_AHP_2018, SRDM_AHP_2021, SRDM_Emken, SRDM_Timon}, and blazar jet electrons~\cite{Blazar_Bhowmick, Blazar_Granelli}, a much lower $m_\chi$ can be searched.

In addition to $\chi$ and target elastic scattering, some branching models, such as neutral current absorption~\cite{Dror_2020, CDEX_DWH_2022} and effective electromagnetic interaction~\cite{PandaX_Nature, NST_Bai}, have increasingly gained attention in DD experiments. Moreover, cosmic rays bombarding the atmosphere are used for $\chi$ detection~\cite{PhysRevD.102.115032, PhysRevLett.123.261802, ARGUELLES2022137363, Flambaum, QiangYuan}. In this work, we investigate a novel $\chi$ source in DD experiments: evaporating primordial black holes (PBHs).

PBH is a possible candidate for DM~\cite{PBH_PRL_251102, PBH_JPG_043001, PBH_MPLA_1440005, PBH_Neutrino_Dasgupta}. Gravitational collapses due to density fluctuations may have resulted in the formation of PBHs soon after the inflationary epoch of the early Universe~\cite{Zel_1967, Harrison_1970, PBHReviewMaxim_2010}. According to general relativity and quantum field theories, when a PBH evaporates, it emits particles, which is well known as Hawking radiation~\cite{Hawking_1974}. The $\chi$ can also be characterized as Hawking radiation~\cite{Morrison_2019, Baldes_2020, Bernal_JCAP03_007, Bernal_PLB_815, Bernal_JCAP03_015, Arbey_2021, Paolo_PRD_095018, Masina_2021, Cheek_015022, Cheek_015023}. An evaporating PBH is a novel source of $\chi$~\cite{Calabrese_L021302, Calabrese_103024, Li_055043, CDEX_ZZH_2023} or other new particles~\cite{Li_095034, Agashe_023014, AxionePBH_JY} beyond the standard model (SM). For convenience, the scenario of ``a PBH evaporating $\chi$" is abbreviated as ``PBHeDM". PBHeDM is model-independent only when considering the gravitational interactions between the dark sector and SM~\cite{Friedlander_2023}; model-dependence comes in when numerical values are derived because of the calculation models.

In this work, PBHeDM is investigated by considering the $\chi$--$e$ elastic scattering in the CDEX-10 experiment~\cite{CDEX_SZ_2020, CDEX_DWH_2022}. The 205.4 kg$\cdot$day exposure data is acquired by p-type point contact germanium detectors~\cite{pPCGe} conducted in the China Jinping Underground Laboratory (CJPL)~\cite{CJPL}.

\section{The $\chi$ emitted from evaporating PBHs}~\label{sec:2}

In 1974, Hawking proposed that evaporating black holes emit matter in the form of thermal radiation~\cite{Hawking_1974}. The relation between the Hawking temperature ($T_{PBH}$) and mass ($M_{PBH}$) of a Schwarzschild black hole is given by the Zeroth law of thermodynamics for black holes based on thermodynamics and the gravitational theory~\cite{Page_1976, Page_1977, MacGibbon}, as described in Eq.~\ref{eq::eq1}. We investigate black holes with a mass of $M_{PBH}\sim\mathcal{O}$(10$^{16}$ g). The mass of these black holes is very small compared to the mass of the Sun; consequently, they are considered primordial rather than cosmological black holes. Thus, we directly refer to these black holes as ``PBHs".

\begin{equation}
	\begin{aligned}
		k_B T_{PBH} = \frac{\hbar c^3}{8\pi GM_{PBH}},
	\end{aligned}
	\label{eq::eq1}
\end{equation}
where $\hbar$ is the reduced Planck constant, $c$ is the speed of light in vacuum, $G$ is the Newtonian constant of gravitation, and $k_B$ is the Boltzmann constant.

A particle $\xi$ emitted from an evaporating PBH can be not only SM particles (such as a photon, a neutrino, or an electron) but also dark radiation. Its emission rate is described in Eq.~\ref{eq::eq2} and can be computed using the publicly available $\tt Blackhawk$-$\tt v2.1$ code~\cite{Arbey_2019, Arbey_2021}.

\begin{equation}
	\begin{aligned}
		\frac{{\rm d}^2N_\xi }{{\rm d}E_\xi {\rm d}t} = \frac{g_\xi }{2\pi } \frac{\Gamma_\xi (E_\xi, M_{PBH}, x)}{exp(E_\xi/k_B T_{PBH})-(-1)^{2s_\xi}},
	\end{aligned}
	\label{eq::eq2}
\end{equation}
where $E_\xi$ is the evaporated $\xi$ energy, $x$ is a set of secondary parameters for the PBH metrics, and $\Gamma_\xi$ is the graybody factor, which represents the probability that a particle generated at the PBH event horizon will escape to infinity in space.

In this work, $\xi = \chi$. Here, we consider the simple case of Dirac fermions ($s_\chi = 1/2$) with four degrees of freedom ($g_\chi=4$) for a chargeless and spinless PBH with a monochromatic distribution $M_{PBH}$ as an example.

\section{The $\chi$ flux reaching the Earth}~\label{sec:3}

The $\chi$ flux emitted from an evaporated PBH in the Milky Way (MW) galaxy and reaches the Earth can be described as follows: 

\begin{equation}
	\begin{aligned}
		\frac{{\rm d}^2\phi_\chi^{MW}}{{\rm d}T_\chi {\rm d}\Omega } = \frac{f_{PBH}}{4\pi M_{PBH}} \int \frac{{\rm d}\Omega_s}{4\pi}\int_{\rm l.o.s.} \rho _{MW}^{\rm NFW}[r(s,\phi)]\frac{d^2N_\chi}{dT_\chi dt}{\rm d}s,
	\end{aligned}
	\label{eq::eq3}
\end{equation}
where the kinetic energy of $\chi$ is $T_\chi = E_\chi - m_\chi$, $\rho_{MW}^{\rm NFW}$ is the DM density of the Milky Way halo in the Navarro-Frenk-White (NFW) profile~\cite{NFW_profile} with a local density $\rho_ \odot$ = 0.4 GeV/cm$^{3}$, $r(s,\phi)$ is the galactocentric distance (the solar distance from the galactic center), $s$ is the line-of-sight (l.o.s.) distance to the PBH, $\phi$ is the angle between the two directions, and $f_{PBH}$ is defined as the fraction of DM composed of PBHs ($f_{PBH} = \rho_{PBH}/\rho_{DM}$).

The constraint results in the scenario of galactic $\gamma$-ray have a slight difference~\cite{COMPTEL} between the Einasto~\cite{Einasto} and NFW~\cite{NFW_profile} profiles. This difference is even smaller after combining the Hawking radiation fluxes from the extragalactic (EG) PBHs.

EDGES 21 cm~\cite{EDGES, EDGES_X} and COMPTEL~\cite{COMPTEL} have imposed the most stringent constraints on $f_{PBH}$ for PBHs with masses ranging from $1 \times 10^{15}$ to $7 \times 10^{16}$ g. The evaporation of PBH heats the original gas and weakens the signal of the 21 cm absorption line. Clark et al.~\cite{EDGES} derive the limit on $f_{PBH}$ from EDGES data. Coogan et al.~\cite{COMPTEL}derive the limit on $f_{PBH}$ from COMPTEL data based on the $\gamma$ evaporation from the galactic PBHs. The $f_{PBH}$ values selected in this work are listed in Table~\ref{tab:addlabel}. A cutoff was applied to $M_{PBH} = 1\times10^{15}$ g since PBHs with $M_{PBH} \lesssim \mathcal{O}(10^{14})$ g evaporate violently and decay away; a cutoff was applied to $M_{PBH} = 7\times10^{16}$ g since the existing experimental constraints on $f_{PBH}$ for large $M_{PBH}$ are close to or larger than 0.1. In addition, there are no valid observational limits on $f_{PBH}$ for $M_{PBH}\sim\mathcal{O}$(10$^{17}$--10$^{23}$) g~\cite{Carr_2021, PBH_JPG_043001}.

\begin{table}[!tbp]
	\centering
	\caption{Selected $f_{PBH}$ values for different $M_{PBH}$s, based
		on the latest constraints obtained from Refs.~\cite{EDGES,COMPTEL}.}
	\begin{tabular*}{\hsize}{@{\quad}@{\extracolsep{\fill}}cccc@{\quad}}
		\hline
		\hline
		$M_{PBH}$ (g) & $f_{PBH}$ & $M_{PBH}$ (g) & $f_{PBH}$ \\
		\hline
		$1\times 10^{15}$ & $1.6\times 10^{-8}$ & $9\times 10^{15}$ & $5.2\times 10^{-5}$ \\
		$2\times 10^{15}$ & $1.8\times 10^{-7}$ & $1\times 10^{16}$ & $7.0\times 10^{-5}$ \\
		$3\times 10^{15}$ & $9.8\times 10^{-7}$ & $2\times 10^{16}$ & $1.0\times 10^{-3}$ \\
		$4\times 10^{15}$ & $3.5\times 10^{-6}$ & $3\times 10^{16}$ & $3.8\times 10^{-3}$ \\
		$5\times 10^{15}$ & $8.6\times 10^{-6}$ & $4\times 10^{16}$ & $7.6\times 10^{-3}$ \\
		$6\times 10^{15}$ & $1.4\times 10^{-5}$ & $5\times 10^{16}$ & $1.4\times 10^{-2}$ \\
		$7\times 10^{15}$ & $2.0\times 10^{-5}$ & $6\times 10^{16}$ & $2.2\times 10^{-2}$ \\
		$8\times 10^{15}$ & $3.5\times 10^{-5}$ & $7\times 10^{16}$ & $3.4\times 10^{-2}$ \\
		\hline
		\hline
	\end{tabular*}
	\raggedright
	\label{tab:addlabel}
\end{table}

The effect of the cosmological red-shift $z(t)$ on the particle kinetic energy should also be taken into account for the EG component:

\begin{equation}
	\begin{aligned}
		\frac{{\rm d}^2\phi_\chi^{EG}}{{\rm d}T_\chi {\rm d}\Omega } =\frac{f_{PBH}\rho _{DM}}{4\pi M_{PBH}} \int ^{t_U} _{t_0} [1+z(t)] \frac{d^2N_\chi }{dT_\chi dt} {\rm d}t,
	\end{aligned}
	\label{eq::eq4}
\end{equation}
where $\rho_{DM}$ = 2.35$\times $10$^{-30}$ g/cm$^3$ is the average DM density of the Universe at the current epoch~\cite{PLANK}, $t_0 = 10^{11}$ s is the Universe time of the matter-radiation equality, and $t_U = 1.38\times10^{10}$ yr is the Universe age; the $\chi$ energy at the source $E_S$ and the $\chi$ energy at the Earth's frame $E_\chi$ obey $1+z(t) = [(E^2_S-m_\chi^2)/(E^2_\chi-m_\chi^2)]^{1/2}$.

\begin{figure}[!tbp]
	\includegraphics[width=0.99\linewidth]{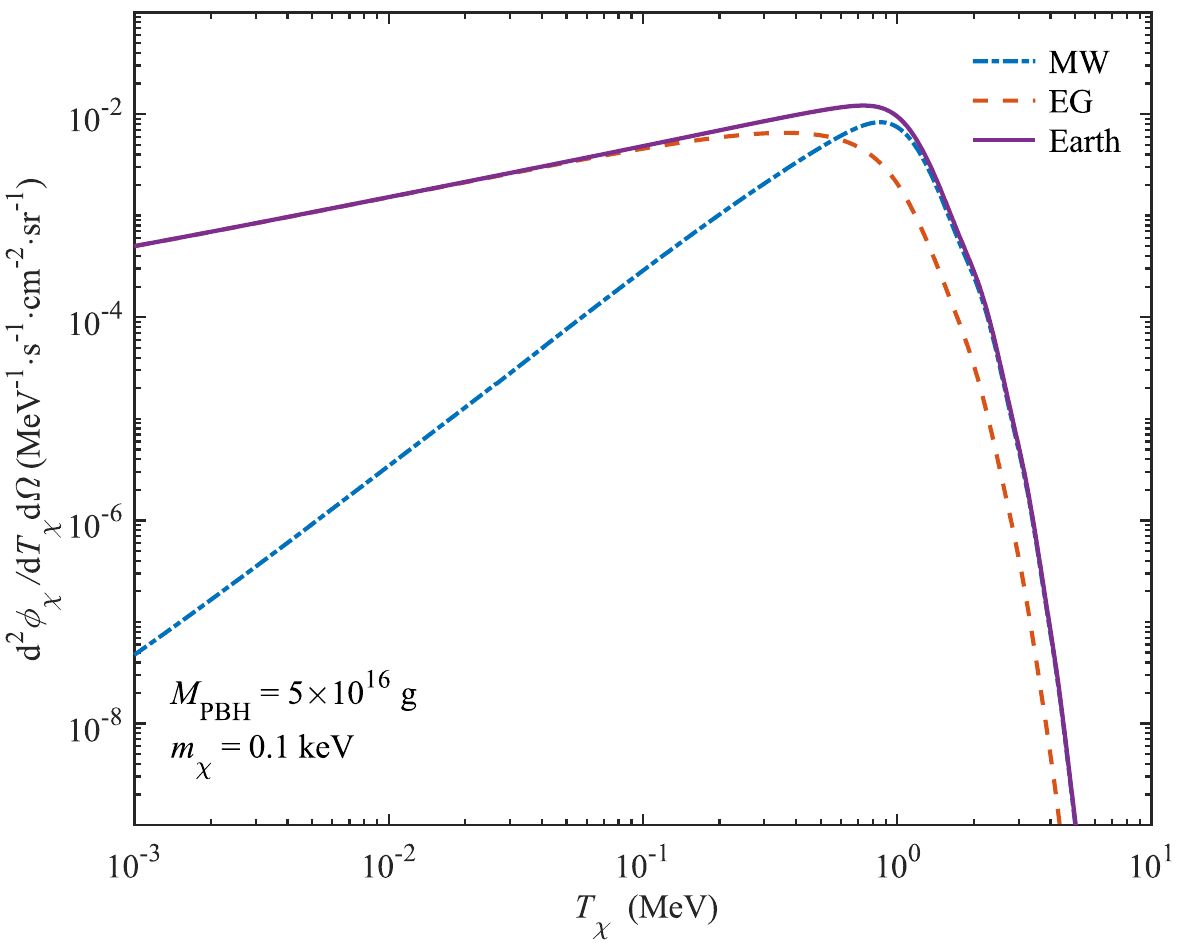}
	\caption{
		Fluxes of $\chi$ with a mass $m_\chi$ = 0.1 keV reaching the Earth from PBHs with masses $M_{PBH}$ = $5\times 10^{16}$ g. The components from galactic (MW) and extragalactic (EG) PBHs are shown by the blue dotted-dashed line and the red dashed line, respectively. $f_{PBH}$s (the fractions of DM composed of PBHs) selected for our calculations are listed in Table~\ref{tab:addlabel}.
	}
	\label{fig::fig1}
\end{figure}

In Eq.~\ref{eq::eq5}, the $\chi$ flux from a PBH to the full sky reaching the Earth has two components: the flux from MW PBHs and the flux from EG PBHs.

\begin{equation}
	\begin{aligned}
		\frac{{\rm d}^2\phi_\chi}{{\rm d}T_\chi {\rm d}\Omega} = \frac{{\rm d}^2\phi_\chi^{MW}}{{\rm d}T_\chi {\rm d}\Omega}+\frac{{\rm d}^2\phi_\chi^{EG}}{{\rm d}T_\chi {\rm d}\Omega},
	\end{aligned}
	\label{eq::eq5}
\end{equation}

Figure~\ref{fig::fig1} shows the flux of $\chi$ with a mass $m_\chi$ = 0.1 keV reaching the Earth from galactic and extragalactic PBHs with masses $M_{PBH}$ = $5\times 10^{16}$ g. Considering the EG component, the cosmological red-shift significantly affects $T_\chi$ and makes a significant contribution to the flux in the low $T_\chi$.

\section{Earth shielding}

When $\chi$ arrive at the Earth, they must travel through the Jinping Mountain before reaching the detector in CJPL. Similar to the target electrons in a detector, the electrons in the rocks may also scatter with $\chi$. For a large scattering cross section with a mean free path in the rocks of less than a few km, $\chi$ will scatter with the electrons in the rocks and lose kinetic energy before reaching the detectors at underground locations. This phenomenon is called ``Earth attenuation'' or ``Earth shielding effect''. We employ a ballistic-trajectory approximation based on Refs.~\cite{CRDM_Bringmann_2019, Calabrese_L021302, Li_055043}. 

The kinetic energy of $\chi$ is reduced by ${\rm d}T$ per length ${\rm d}l$ due to its isotropic elastic scattering caused by the electrons in the medium. After traveling a distance $d$, the kinetic energy of $\chi$ is reduced from $T_0$ to $T_d$, and the flux of $\chi$ changes from $\frac{{\rm d}^2\phi_\chi}{{\rm d}T_0 {\rm d}\Omega }$ to $\frac{{\rm d}^2\phi_\chi^d}{{\rm d}T_d {\rm d}\Omega }$, where
\begin{equation}
	\begin{aligned}
		\frac{{\rm d}^2\phi_\chi^d}{{\rm d}T_d {\rm d}\Omega } \approx \frac{4 m_\chi^2e^\tau}{(2 m_\chi+T_d-T_de^\tau)^2} \left(\frac{{\rm d}^2\phi_\chi}{{\rm d}T_0 {\rm d}\Omega }\right),
	\end{aligned}
	\label{eq::eq6}
\end{equation}
with
\begin{equation}
	\begin{aligned}
		T_d(T_0) = \frac{2 m_\chi T_0}{2 m_\chi e^\tau - T_0(1+e^\tau)},
	\end{aligned}
	\label{eq::eq7}
\end{equation}
where $\tau = d/l$, $d$ = 2400 m is the rock overburden depth of the CJPL~\cite{CJPL}, and $l$ is the interaction length given as follows:

\begin{equation}
	\begin{aligned}
		l=\left[\Bar n_e \sigma_{\chi e} \frac{2 m_e m_\chi}{(m_e+m_\chi)^2} \right]^{-1},
	\end{aligned}
	\label{eq::eq8}
\end{equation}
where $m_e$ is the electron mass and $\sigma_{\chi e}$ is the $\chi$--electron elastic-scattering cross section. The average number density of the electrons in the rocks ($\Bar n_e = 8.1\times 10^{23}$ cm$^{-3}$) is computed from the density and the elemental composition of the rocks, as described in Ref.~\cite{CDEX_LZZ_2022}.

\section{Expected spectra in the detector}

When $\chi$ arrives at the detector, there is a probability of elastic scattering and leaving some recoil energy of $E_r$. If $E_r$ is higher than the binding energy of a shell electron, the event may be recorded by the detector. For $\chi$, the number of target electrons is $\left(N_{Ge} Z^{Ge}_{eff} \sigma_{\chi e}\right)$, where $N_{Ge}$ = 8.30$\times$10$^{21}$ g$^{-1}$ is the number density of germanium atoms and $Z^{Ge}_{eff}$ is the effective electron number of germanium listed in Table~\ref{tab:addlabel2}~\cite{GeEnergy1, GeEnergy2}. For instance, when $E_r =$ 1 keV in a scattering, then 22 out of 32 shell electrons of the target atom are considered target candidates by $\chi$, and the effective electron number is 22. Considering once again that the scattering is isotropic, the differential event rate of the $\chi$-electron scattering in the detector is given in Eq.~\ref{eq::eq9}~\cite{LZ_neutrino, Li_055043}.

\begin{equation}
	\begin{aligned}
		\frac{{\rm d}R_r}{{\rm d}E_r}(E_r) = N_{Ge} Z^{Ge}_{eff} \sigma_{\chi e}\int \frac{{\rm d}^2\phi_\chi^d}{{\rm d}T_d {\rm d}\Omega} \frac{\Theta(E_r^{max}-E_r)}{E_r^{max}} {\rm d}T_d {\rm d}\Omega,
	\end{aligned}
	\label{eq::eq9}
\end{equation}
where $\Theta$ is the Heaviside function and $E_r^{max}$ is the maximum value of the recoil energy $E_r$ limited by 
\begin{equation}
	\begin{aligned}
		E_r^{max} = \frac{T_d^2+2 m_\chi T_d}{T_d+(m_\chi +m_e)^2/(2 m_e)}.
	\end{aligned}
	\label{eq::eq10}
\end{equation}

When estimating the average scattering cross section $\Bar \sigma_{\chi e}$ according to Ref.~\cite{Calabrese_103024}, calculation is difficult as germanium crystal has a band structure and Ge atoms cannot be considered isolated~\cite{Griffin, CDEX_ZZY_2022}. Moreover, most $\chi$ are relativistic particles. Therefore, a more elaborate approach should be considered in future calculations.

\begin{figure}[!tbp]
	\includegraphics[width=0.99\linewidth]{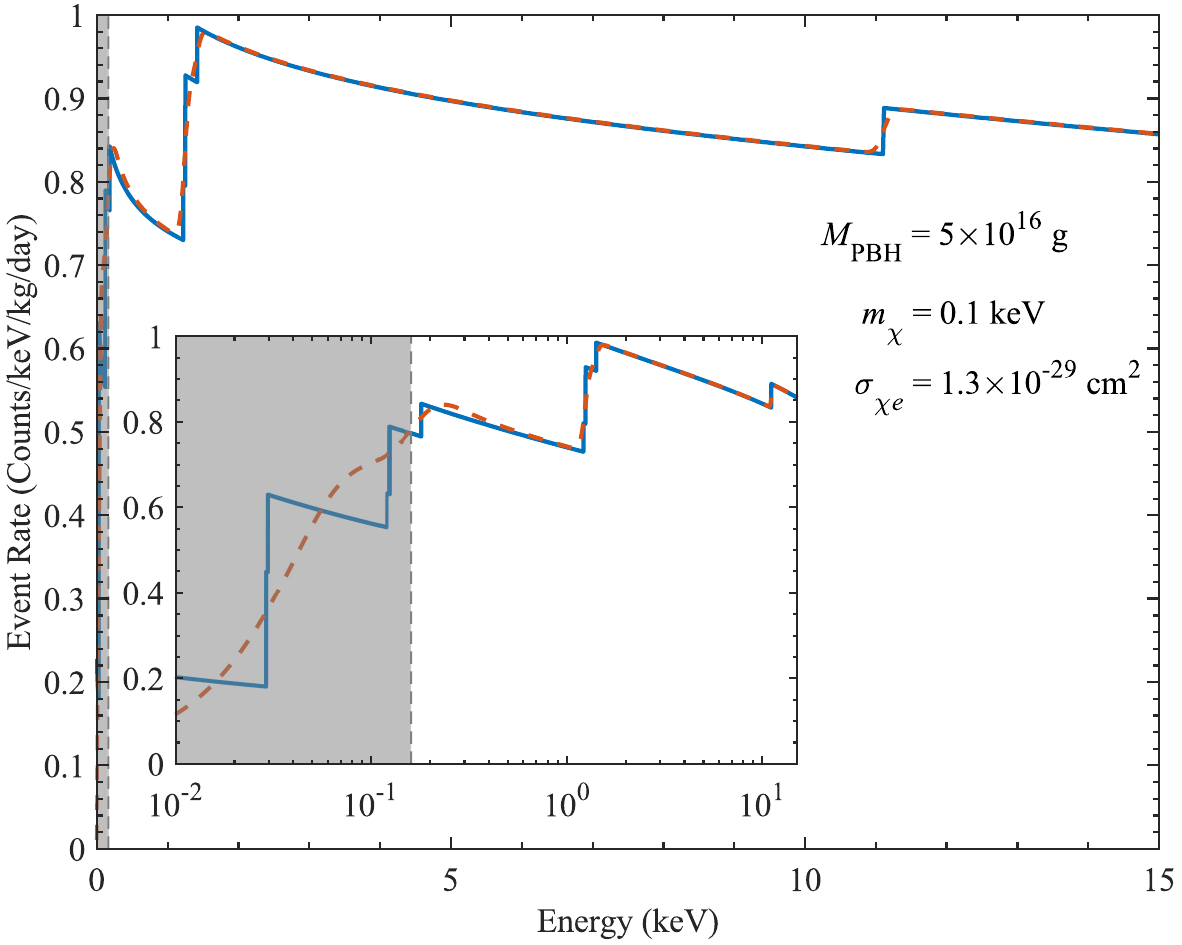}
	\caption{
		Recoil energy (blue solid line) and expected energy (red dashed line) event rates for $M_{PBH} = 5\times10^{16}$ g, $m_{\chi} = 0.1$ keV, and $\sigma_{\chi e} = 1.3\times10^{-29}$ cm$^2$. The energy resolution in the CDEX-10 experiment is $\sigma_{Res}(E) = 35.8+16.6\cdot E^{1/2}$ eV, where $E$ is in keV units~\cite{CDEX_JH_2019, CDEX_ZZY_2022}. The fine black dashed line indicates the 160-eV analysis threshold.
	}
	\label{fig::fig2}
\end{figure}

By convoluting the energy resolution obtained from the calibration, the recoil event rate $\frac{{\rm d}R_r}{{\rm d}E_r}$ is converted into the expected event rate $\frac{{\rm d}R_e}{{\rm d}E_e}$ recorded by the CDEX-10 detector. The energy resolution in the CDEX-10 experiment is $\sigma_{Res}(E) = 35.8+16.6\cdot E^{1/2}$ eV, where $E$ is in keV units~\cite{CDEX_JH_2019,CDEX_ZZY_2022}. The spectra before (blue solid line) and after (red dashed line) considering the energy resolution are shown in Fig.~\ref{fig::fig2}. There are some step-like fine spectral structures with the binding energy of Ge atom, which are retained with the excellent energy resolution of the CDEX-10 detector, although somewhat smoothed.

\begin{table}[!tbp]
	\centering
	\caption{Effective electron number of germanium $Z^{Ge}_{eff}$~\cite{GeEnergy1, GeEnergy2}.}
	\begin{tabular*}{0.7\hsize}{cc}
		\hline
		\hline
		~\quad~\quad $Z^{Ge}_{eff}$~\quad~\quad & ~\quad$E_r$ (keV)~\quad \\
		\hline
		32 & 1.11e1 $<E_r$             \\
		30 & 1.41e0 $<E_r\leq$ 1.11e1  \\
		28 & 1.25e0 $<E_r\leq$ 1.41e0  \\
		24 & 1.22e0 $<E_r\leq$ 1.25e0  \\
		22 & 1.80e-1$<E_r\leq$ 1.22e0  \\
		20 & 1.25e-1$<E_r\leq$ 1.80e-1 \\
		16 & 1.21e-1$<E_r\leq$ 1.25e-1  \\
		14 & 2.98e-2$<E_r\leq$ 1.21e-1 \\
		10 & 2.92e-2$<E_r\leq$ 2.98e-2 \\
		4  & 7.20e-3$<E_r\leq$ 2.92e-2 \\
		0  & $E_r\leq$ 7.20e-3 \\
		\hline
		\hline
	\end{tabular*}
	\label{tab:addlabel2}
\end{table}

\section{Results}

In this work, the physical analysis is based on the CDEX-10's 205.4 kg$\cdot$day exposure data in the 0.16--12.06 keV range. There are four steps in the preprocessing procedures for the exposure data~\cite{CDEX_JH_2018, CDEX_JH_2019}:

1) The zero peak (random trigger events), the 8.98 keV peak ($^{65}$Zn), the 10.37 keV peak ($^{68}$Ge) were used to calibrate the energy in the low-energy region.

2) The events with anomalous electronic noise profiles were rejected in the pedestal cut.

3) Based on the linear relationship between the integral value and the maximum value of the physical event pulse, the physicals signal can be discriminated from the electronic noises near the energy threshold.

4) Events depositing energy in the surface layer of the detector generate a slow rising pulse and an incomplete charge collection. The energy measured in these events are low and should be removed.

\begin{figure}[!tbp]
	\includegraphics[width=0.99\linewidth]{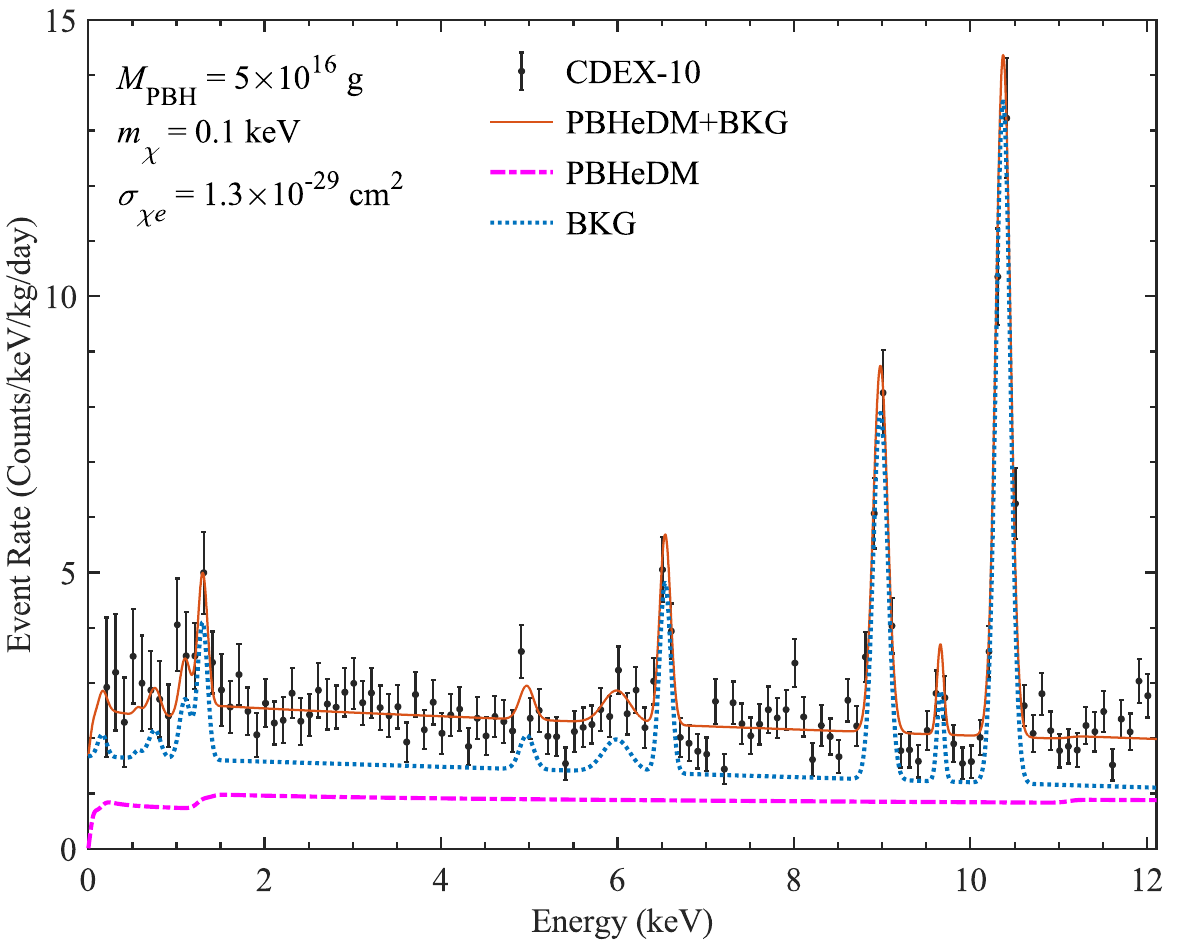}
	\caption{
		Measured spectrum using 205.4 kg$\cdot$day exposure data in the 0.16--12.06 keV range obtained from the CDEX-10 experiment~\cite{CDEX_SZ_2020, CDEX_DWH_2022} (black points with error bars). The red solid line shows the best fit of the minimum-$\chi^2$ for $M_{PBH}$ = $5\times10^{16}$ g. The pink dashed-dotted line and the blue dotted line show the ``PBHeDM" and ``BKG" components, respectively.
	}
	\label{fig::fig3}
\end{figure}

The measured spectrum is shown by the black points with error bars in Fig.~\ref{fig::fig3}. Then, a minimum-$\chi^2$ analysis, as described in Eq.~\ref{eq::eq11}, is performed. The expected event rate $\frac{{\rm d}R_e}{{\rm d}E_e}$, denoted by $S$, is a function of $\left(M_{PBH}, f_{PBH}, m_{\chi}, \sigma_{\chi e}\right)$. The background $BKG$ depends on $\left(a, b, I_K, \sigma_K, \sigma_L,\sigma_M \right)$ following Eq.~\ref{eq::eq12}. The best estimate of $BKG$ can be evaluated by scanning this parameter space in the minimum-$\chi^2$ analysis, from which $S$ can be derived. The components of the best fit result are plotted in Fig.~\ref{fig::fig3}.

\begin{table}[!tbp]
	\centering
	\caption{Identified X-ray peaks in the CDEX-10 measured spectrum.}
	\begin{tabular*}{\hsize}{@{\quad}@{\extracolsep{\fill}}cccccc@{\quad}}
		\hline
		\hline
		$E$ (keV) & Nuclide & Shell & $E$ (keV) & Nuclide & Shell\\
		\hline
		0.14 & $^{68}$Ga & $M$ & 1.30 & $^{68}$Ge & $L$\\
		0.16 & $^{68}$Ge & $M$ & 4.97 & $^{49}$V  & $K$\\
		0.56 & $^{49}$V  & $L$ & 5.99 & $^{54}$Mn & $K$\\
		0.70 & $^{54}$Mn & $L$ & 6.54 & $^{55}$Fe & $K$\\
		0.77 & $^{55}$Fe & $L$ & 8.98 & $^{65}$Zn & $K$\\
		1.10 & $^{65}$Zn & $L$ & 9.66 & $^{68}$Ga & $K$\\
		1.19 & $^{68}$Ga & $L$ & 10.4 & $^{68}$Ge & $K$\\
		\hline
		\hline
	\end{tabular*}
	\raggedright
	\label{tab:addlabel3}
\end{table}

\begin{equation}
	\begin{aligned}
		\chi^2 &\left(M_{PBH}, f_{PBH}, m_{\chi}, \sigma_{\chi e}\right) = \\
		& \sum_i \frac{\left[n_i-BKG_i-S_i\left(M_{PBH}, f_{PBH}, m_{\chi}, \sigma_{\chi e}\right)\right]^2}{\Delta^2_i},
	\end{aligned}
	\label{eq::eq11}
\end{equation}
where $n_i$ and $\Delta _i$ are the measured data and the standard deviation of the statistical and systematic components at the $i$-th energy bin, respectively.

\begin{equation}
	\begin{aligned}
		BKG = \left(a\cdot E+b\right)&+&\sum_M\frac{I_M}{\sqrt{2\pi}\sigma_M}exp{\left[-\frac{(E-E_M)^2}{2\sigma_M^2}\right]}\\
		&+&\sum_L\frac{I_L}{\sqrt{2\pi}\sigma_L}exp{\left[-\frac{(E-E_L)^2}{2\sigma_L^2}\right]}\\
		&+&\sum_K\frac{I_K}{\sqrt{2\pi}\sigma_K}exp{\left[-\frac{(E-E_K)^2}{2\sigma_K^2}\right]},\\
	\end{aligned}
	\label{eq::eq12}
\end{equation}
where $\left(a\cdot E+b\right)$ is the linear platform; $I_M$s, $I_L$s, and $I_K$s are the intensities of the $M$, $L$, and $K$-shell X ray peaks, respectively; $\sigma_M$s, $\sigma_L$s, and $\sigma_K$s are the energy resolutions of the $M$, $L$, and $K$-shell X-ray peaks, respectively. Referring to our previous studies~\cite{CRDM_CDEX_2022, CDEX_DWH_2022}, fourteen X-ray peaks were identified and listed in Table~\ref{tab:addlabel3}. $I_K$s are used to constrain $I_M$s and $I_L$s with known $K$/$L$ and $K$/$M$ ratios~\cite{LKRatio, CDEX_JH_2018}.

\begin{figure}[!tbp]
	\includegraphics[width=0.99\linewidth]{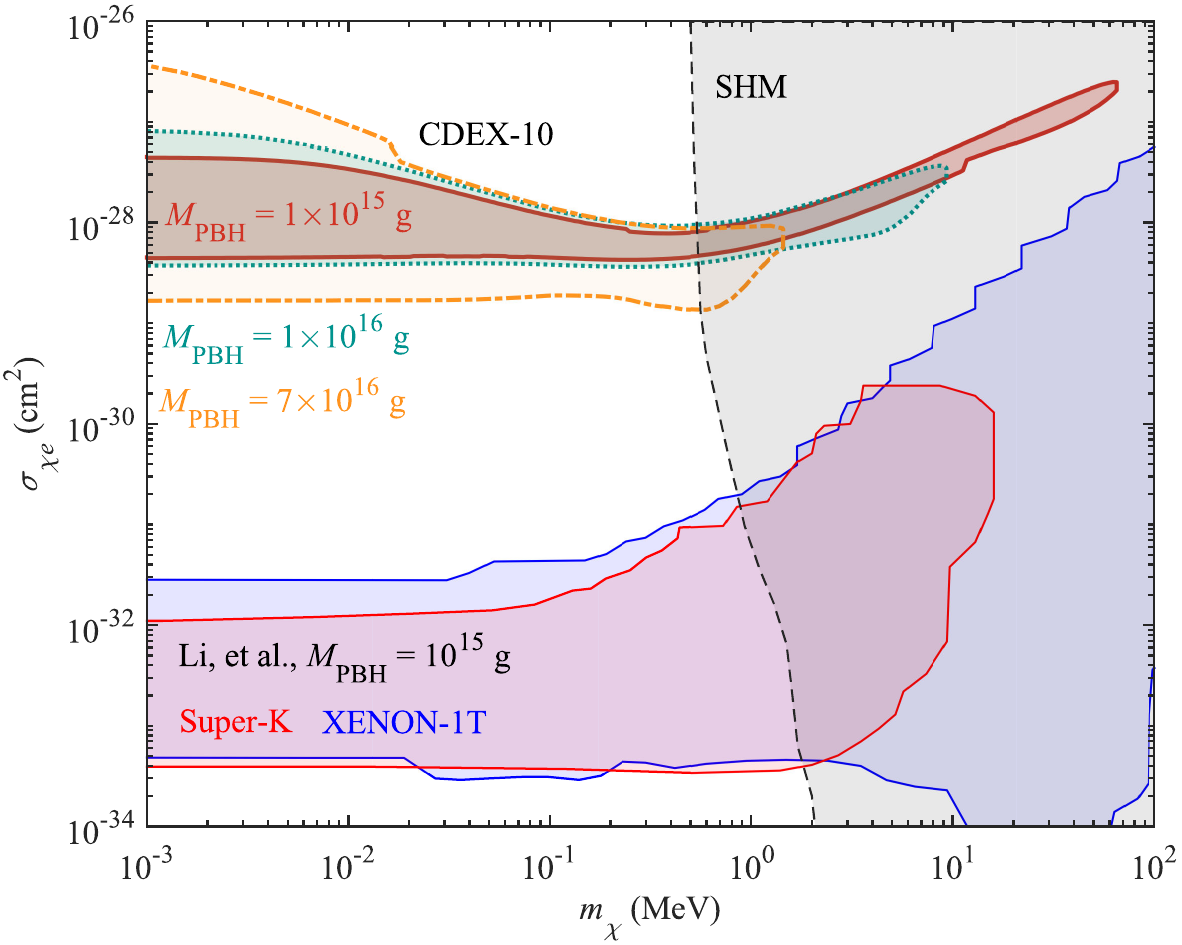}
	\caption{
		Exclusion regions ($m_\chi$, $\sigma_{\chi e}$) in PBHeDM with a 90\% CL for $M_{PBH}$ = (1, 10, 70)$\times10^{15}$ g, and $f_{PBH} = 1.6\times10^{-8}, 7.0\times10^{-5}, 3.4\times10^{-10}$ respectively as listed in Table~\ref{tab:addlabel}. For comparison, the limit on the cold $\chi$ in the SHM~\cite{Chi-e_SENSEI_2020, Chi-e_DAMIC_2019, Chi-e_EDELWEISS_2020, Chi-e_SuperCDMS_2019, Chi-e_DarkSide_2018, Chi-e_XENON_2019, Chi-e_Panda_2021} is also shown in dashed gray shadow. The results obtained from the phenomenological interpretations of Li et al.~\cite{Li_055043} for the Super-Kamiokande and XENON-1T data for $M_{PBH} = 10^{15}$ g and $f_{PBH} = 3.9\times10^{-7}$ are plot respectively in red and blue. For visual clarity, the results from Calabrese et al.~\cite{Calabrese_103024} are omitted from the figure.
	}
	\label{fig::fig4}
\end{figure}

There is no significant signal characteristic in the measured spectrum using CDEX-10's 205.4 kg$\cdot$day exposure data in the 0.16--12.06 keV range. Thus, the 90\% confidence level (CL) exclusion regions ($m_\chi$, $\sigma_{\chi e}$) for $M_{PBH}$ = (1 -- 70)$\times10^{15}$ g can be derived using the Feldman--Cousins unified approach with $\Delta\chi^2$ = 1.64~\cite{FCChiSquare}. The constraints on PBHeDM for $M_{PBH}$ = (1, 10, 70)$\times10^{15}$ g are selected and plotted in Fig.~\ref{fig::fig4}. The limit on the cold $\chi$ in the SHM~\cite{Chi-e_SENSEI_2020, Chi-e_DAMIC_2019, Chi-e_EDELWEISS_2020, Chi-e_SuperCDMS_2019, Chi-e_DarkSide_2018, Chi-e_XENON_2019, Chi-e_Panda_2021} is also shown in dashed gray shadow.

In the new concept of PBHeDM, the parameter space of ($m_\chi$, $\sigma_{\chi e}$) can be effectively searched in the low $m_\chi$. The results obtained from the phenomenological interpretations of Li et al.~\cite{Li_055043} for the Super-Kamiokande and XENON-1T data for $M_{PBH} = 10^{15}$ g and $f_{PBH} = 3.9\times10^{-7}$ are plot respectively in red and blue.  In Calabrese et al.'s work~\cite{Calabrese_103024}, the atom ionization with an outgoing nonrelativistic free electron caused by the $\chi$ -- $e$ scattering is treated by the theory of nondenormalized effective fields, and then the lower limits on the mean cross section $\bar{\sigma}_{\chi e}$ are derived from the Super-Kamiokande and XENON-1T data. For visual clarity, the results are omitted from the figure. CDEX-10's radiation background is about three orders of magnitude higher than XENON-1T's. But CDEX-10's energy threshold (0.16 keV) is lower than XENON-1T's (1 keV). Thus, CDEX-10 excludes the space above them.

\begin{figure}[!tbp]
	\includegraphics[width=0.99\linewidth]{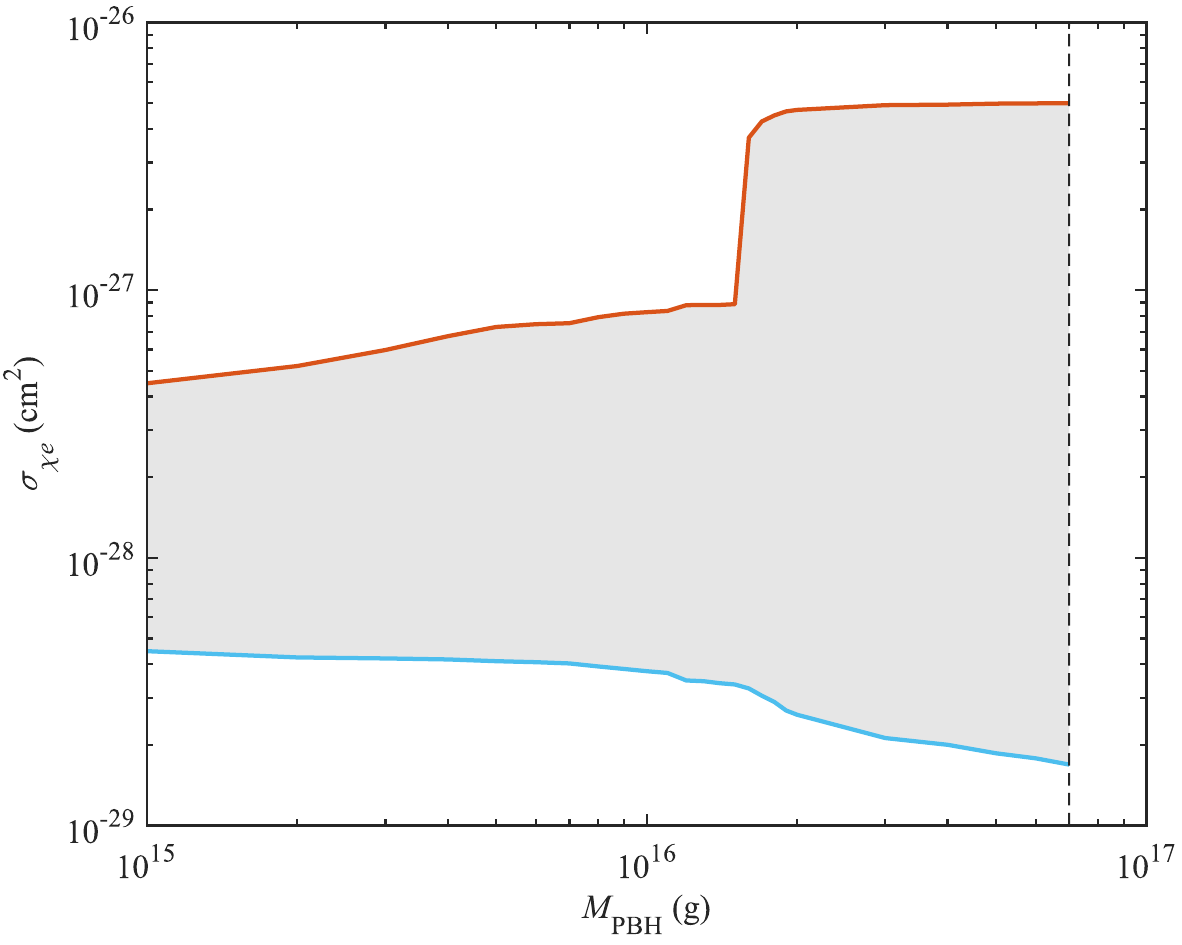}
	\caption{
		Numerical exclusion regions ($M_{PBH}$, $\sigma_{\chi e}$) with a 90\% CL for $m_{\chi}$ $\lesssim$ 0.1 keV. The lower and upper bounds on $\sigma_{\chi e}$ are shown by the blue and red solid lines, respectively. The gray-shaded parameter space in between is probed and excluded. 
	}
	\label{fig::fig5}
\end{figure}

As a supplement, the projection of ($M_{PBH}$, $f_{PBH}$, $m_\chi$ = 0.1 keV, $\sigma_{\chi e}$) on ($M_{PBH}$, $\sigma_{\chi e}$) is drawn in Fig.~\ref{fig::fig5} numerically but not physically because both the upper- and lower-bound lines of the exclusion regions ($m_{\chi}$, $\sigma_{\chi e}$) for $m_{\chi}$ $\lesssim$ 0.1 keV are flat.

\section{Discussion}

To calculate the physical ($m_{\chi}$, $\sigma_{\chi e}$) and numerical ($M_{PBH}$, $\sigma_{\chi e}$) results, the $f_{PBH}\left(M_{PBH}\right)$ values listed in Table~\ref{tab:addlabel} used as inputs in Eq.~\ref{eq::eq3} and~\ref{eq::eq4}. The flux reaching the Earth of $\frac{{\rm d}^2\phi_\chi}{{\rm d}T_\chi {\rm d}\Omega}$ in Eq.~\ref{eq::eq5} is proportional to the $f_{PBH}$. The red line in Fig.~\ref{fig::fig5} has  a step-wise structure at $M_{PBH} \simeq 1.5\times10^{16}$ g because it is required the stronger Earth attenuation to reduce the flux reaching the detector. However, the effect of Earth shielding on the lower limit is not significant and the $\frac{{\rm d}^2\phi_\chi^d}{{\rm d}T_d {\rm d}\Omega}$ in Eq.~\ref{eq::eq6} is approximately equal to the $\frac{{\rm d}^2\phi_\chi}{{\rm d}T_\chi {\rm d}\Omega}$ in Eq.~\ref{eq::eq5}. The blue line does not vary as rapidly as the red line because the expected spectrum exhibits some fine shapes.

\begin{figure}[!tbp]
	\includegraphics[width=0.99\linewidth]{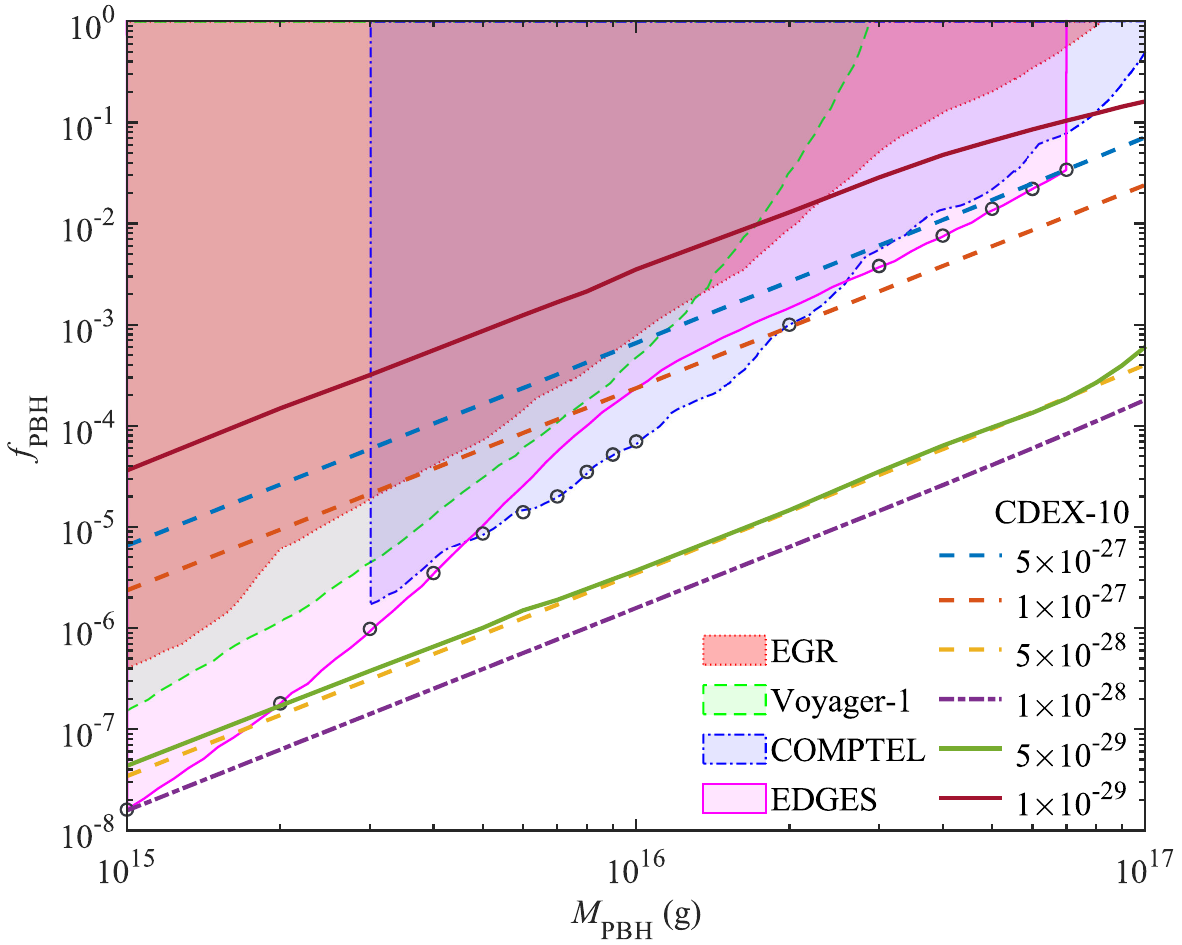}
	\caption{
		Upper bounds with a 90\% CL on $f_{PBH}\left(M_{PBH}\right)$ for different $\sigma_{\chi e}$s ($m_\chi$ $\lesssim$ 0.1 keV). The numbers $5\times10^{-27}$, $1\times10^{-27}$, and so on, shown in the legend are in cm$^2$ and correspond to $\sigma_{\chi e}$s.  Some strong existing bounds obtained from the extragalactic $\gamma$-ray background (EGR)~\cite{EGR}, EDGES 21 cm~\cite{EDGES}, Voyager-1~\cite{Voyager}, and COMPTEL~\cite{COMPTEL} observations are shown; the strongest bounds marked with black circles correspond to $f_{PBH}$ listed in Table~\ref{tab:addlabel} and were selected as inputs in Eqs.~\ref{eq::eq4} and~\ref{eq::eq5}. The results obtained from the phenomenological interpretations of Calabrese et al.~\cite{Calabrese_103024} and Li et al.~\cite{Li_055043} for the Super-Kamiokande and XENON-1T data are omitted from the figure for visual clarity.
	}
	\label{fig::fig6}
\end{figure}

There remains a degenercy in $\sigma_{\chi e}(m_{\chi})$ and $f_{PBH}(M_{PBH})$ from this work. If ($m_{\chi}, \sigma_{\chi e}$) can be determined in the future in other scenarios, such as SHM, we can search for evaporating PBHs through $\chi$ as particles in the SM such as $\gamma$ photons. Here, $\sigma_{\chi e}$ in turn becomes the input value and $f_{PBH}$ becomes the experimental output limit in DD experiments. For $m_\chi \lesssim 0.1$ keV, upper bounds with a 90\% CL on $f_{PBH}\left(M_{PBH}\right)$ for different $\sigma_{\chi e}$s from CDEX-10 are plotted in lines in Fig.~\ref{fig::fig6}.

All of the $f_{PBH}$s (including those from COMPTEL from other experiments) increase as $M_{PBH}$ increase because of the decreased Hawking temperature. For larger $M_{PBH}$, a lower energy threshold in the analysis is required. Due to the effect of Earth shielding, $f_{PBH}$ decreases with the decrease in $\sigma_{\chi e}$ between $\mathcal{O}(5\times10^{-27})$ and $\mathcal{O}(1\times10^{-28})$ cm$^2$ in Fig.~\ref{fig::fig6}. Therefore, an accurate Earth shielding model rather than an approximation is required. This remains an open research area that will be the theme of our future efforts. Between $\mathcal{O}(1\times10^{-28})$ and $\mathcal{O}(1\times10^{-29})$ cm$^2$, $f_{PBH}$ increases with the decrease in $\sigma_{\chi e}$.

It is worth noting that in Fig.~\ref{fig::fig6}, the CDEX-10 curve rises more slowly than the black dots (which is the current best $f_{PBH}$ limit obtained from cosmological observations~\cite{EGR, EDGES, Voyager, COMPTEL}). The XENON-1T's results shown in Fig.~\ref{fig::fig3} in Ref.~\cite{Calabrese_103024} and in Fig.~\ref{fig::fig4} in Ref.~\cite{Li_095034} exhibit the same trend. For example, the curve of ($m_{\chi} =$ 0.1 keV, $\sigma_{\chi e} = 1\times10^{-29}$ cm$^2$) is out of the shaded part when $M_{PBH}\gtrsim 8\times10^{16}$ g. This indicates that PBHeDM could be used to search for evaporating PBHs with large $M_{PBH}$s and imposes constraints on $f_{PBH}$s. DD experiments with low radiation background and low energy threshold, such as XENON and CDEX, are expected to fill the gap in ($M_{PBH}$, $f_{PBH}$) for $M_{PBH}\sim \mathcal{O}(10^{18})$ g in Ref.~\cite{Carr_2021, PBH_JPG_043001} if the physical properties of $\chi$ can be clarified in the future.

\section{Summary}

PBHeDM is a novel non-thermal DM particle. In the presence of ($M_{PBH}$, $f_{PBH}$) compatible with present bounds, CDEX-10 experiment excludes $\sigma_{\chi e} \sim \mathcal{O}(10^{-28})$ cm$^2$ for $\chi$ with a mass $m_{\chi}\lesssim$ 0.1 keV for $M_{PBH}$ ranging from $1\times10^{15}$ g to $7\times10^{16}$ g. This is a parameter space that previous works have not explored. In the future, if $\sigma_{\chi e}\left( m_{\chi} \right) $ is known, then DD experiments will have the potential to give limits on $f_{PBH}\left( M_{PBH}\right)$ for the larger $M_{PBH}$.

\section*{Acknowledgements}

This work was supported by the National Key Research and Development Program of China (Grants No. 2023YFA1607100 and No. 2022YFA1605000) and the National Natural Science Foundation of China (Grants No. 12322511, No. 12175112, No. 12005111, and No. 11725522). We acknowledge the Center of High performance computing, Tsinghua University for providing the facility support. We would like to thank CJPL and its staff for hosting and supporting the CDEX project. CJPL is jointly operated by Tsinghua University and Yalong River Hydropower Development Company.

\bibliography{ePBH_DM.bib}
\end{document}